\documentclass[%
 reprint,
superscriptaddress,     
 amsmath,amssymb,
 aps,
]{revtex4-2}

\usepackage{graphicx}
\usepackage{dcolumn}
\usepackage{bm}
\usepackage{balance}

\usepackage{float} 

\begin{document}

\preprint{APS/123-QED}

\title{Nonreciprocal ENZ-Dielectric Bilayers:\\ Enhancement of Nonreciprocity from a Nonlinear Transparent Conducting Oxide Thin Film at Epsilon-Near-Zero (ENZ) Frequency}


\author{Diego M. Sol\'{i}s}
\affiliation{Department of Electrical and Systems Engineering, University of Pennsylvania, Philadelphia, Pennsylvania, USA}
\author{Nader Engheta}
\affiliation{Department of Electrical and Systems Engineering, University of Pennsylvania, Philadelphia, Pennsylvania, USA}

\date{\today}

\begin{abstract}
We envision the use of an indium tin oxide (ITO) thin film as part of a bi-layered silicon-photonics subwavelength device to boost nonlinearity-assisted all-passive nonreciprocal behavior. The asymmetric $p$-polarized oblique excitation of a mode near the epsilon-near-zero (ENZ) frequency, with highly-confined and enhanced normal electric field component and large absorption, allows to harness ITO's strong ultrafast nonlinear response for the generation of a notable nonreciprocal performance in the two-port element. Though limited by loss, we find the device's optimal operational point and the maximum nonreciprocal transmittance ratio attainable vs. light intensity---including an apparent upper bound slightly over 2---, and we perform exhaustive numerical simulations considering nonlinear processes of both anharmonic and thermal nature that validate our predictions, including steady-state and pulsed-laser excitations. 
\end{abstract}

\maketitle


\section{\label{sec:level1}Introduction}
The nonreciprocal propagation of light (implying e.g. that, in a two-port system, transmission is different when exciting it from opposite sides) is a key ingredient in the pursue of all-optical nanocircuits. Lorentz reciprocity can be broken with (i) some form of time-reversal antisymmetric biasing \cite{RevModPhys.17.343} (be it magneto-optical \cite{Wang2009,5565359,doi:10.1021/nl900007k} or, in the sub-terahertz regime, a transistor biasing network \cite{doi:10.1063/1.3615688,Wang13194}), (ii) spatially-inhomogeneous temporal modulation \cite{Yu2009,PhysRevLett.109.033901,doi:10.1021/ph400058y,Tzuang2014,Shen2016} or (iii) nonlinear optical transitions \cite{Soljacic:03,Fan447,Shi2015,Mahmoud2015,8445668,Sounas2018,PhysRevLett.120.087401,9446545,9445734}, the latter being the only one allowing for an all-passive approach (though these can also be leveraged to pursue nonreciprocal transmission contrast with e.g. $\mathcal{P}\mathcal{T}$-symmetric (active) systems \cite{PhysRevA.82.043803,Chang2014}). Its operation principle consists of bringing together geometrical asymmetry and some sort of nonlinearity (typically Kerr-like): given that the field distribution inside such asymmetric structure varies with the excitation port, so does the nonlinear response and thus transmission. In this letter, we propose to boost this nonlinear mechanism through the use of an indium tin oxide (ITO) thin layer at its Epsilon-Near-Zero (ENZ) frequency as part of a Silicon-ITO two-port device. We show that, when such structure is resonantly excited under oblique $p$-polarized light, the strong (nonresonant) ENZ nonlinear response dramatically increases nonreciprocal transmission.

\section{Proof of Concept}
\begin{figure}[t]
\includegraphics[width=3.4in]{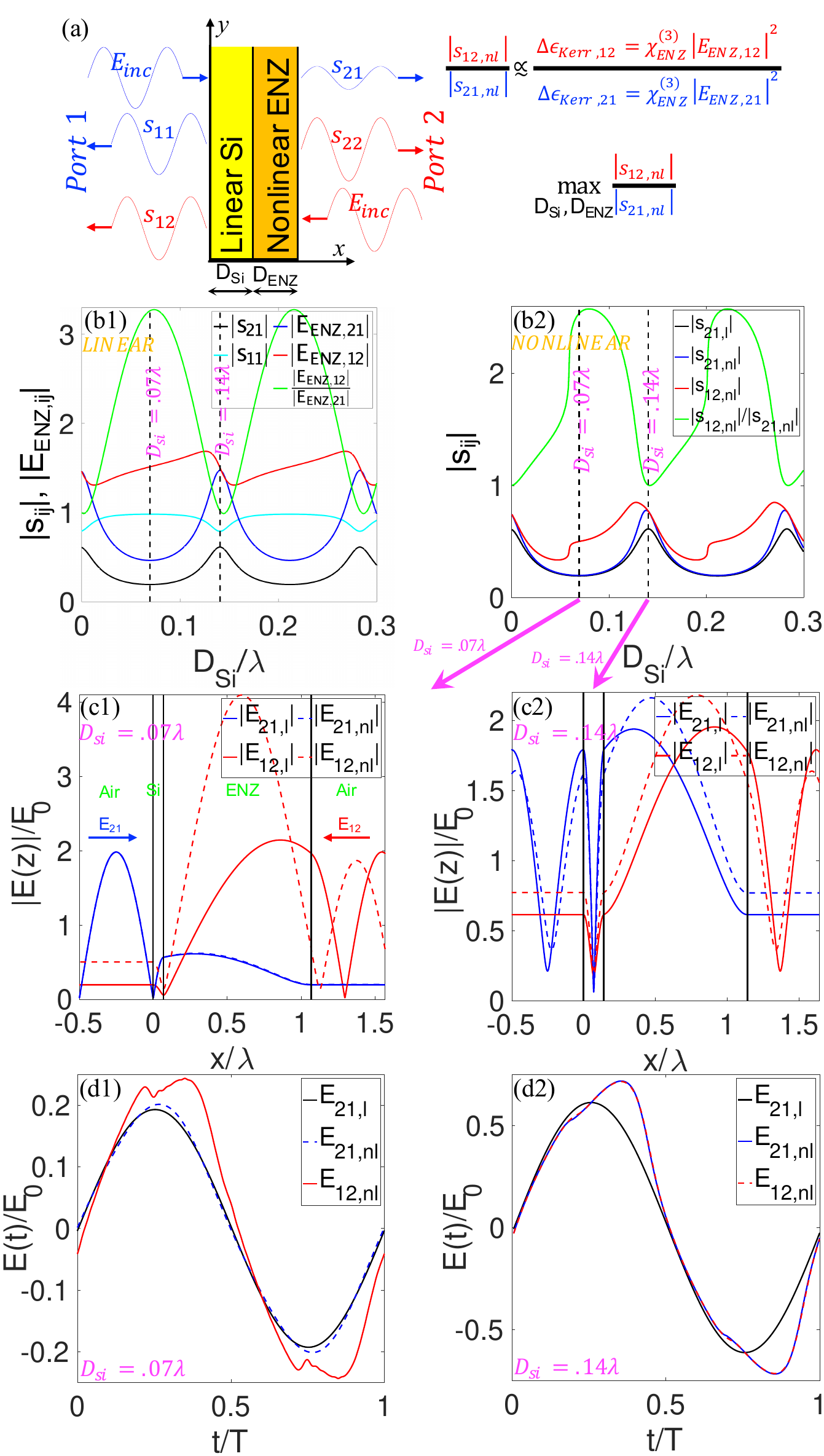}
\caption{Toy model of nondispersive nonreciprocal bilayer device. (a) Schematic plot of the asymmetric nonlinear bilayer two-port device. (b1) $s$-parameters and $\mathbf{x}$-averaged electric field magnitude inside the ENZ slab as a function of the silicon layer's thickness $D_{Si}$ in the linear problem. (b2) Transmission coefficients vs $D_{Si}$ in the nonlinear nonreciprocal problem ($\chi^{(3)}\!=\!3.33\!\times\!10^{-21}\text{(V/m)}^{-2}$, $E_0\!=10^9\text{(V/m)}$). (c1),(c2) Magnitude of electric field phasor vs $x$ in the monochromatic nonlinear problem when $D_{Si}$ minimizes (c1) and maximizes (c2) $\vert s_{21}\vert$, respectively. (d1),(d2) Electric field at the output plane vs. time in the harmonic-generation nonlinear problem.}
\label{fig:1}
\end{figure}

\newcommand{\comment}[1]{}
Our proposed idea is schematically simplified in Fig.~\ref{fig:1}a for a Silicon-ENZ bilayer infinitely extended in the transverse $\mathbf{y}\mathbf{z}$-directions under normal incidence: the structural asymmetry brought about by the two stacks gives rise to different electric-field strengths inside the nonlinear layer when exciting port 1 or 2, and thus different strengths of the nonlinear response that, in turn, alter the transmission coefficients differently. Now, it turns out that time-reversal symmetry dictates that, for any linear structure, the field-distribution asymmetry for excitation from opposite ports is inversely proportional to transmission \cite{Mahmoud2015,8445668}; as pointed out in \cite{PhysRevLett.118.154302}, the ratio of field distributions $\frac{\vert\mathbf{E}_{12}\vert^2}{\vert\mathbf{E}_{21}\vert^2}$ for an arbitrary two-port lossless linear network is bounded as $\frac{1-R}{1+R}\!\leq\!\frac{\vert\mathbf{E}_{12}\vert^2}{\vert\mathbf{E}_{21}\vert^2}\!\leq\!\frac{1+R}{1-R}$, with $R\!=\!\vert s_{11}\vert\!=\!\vert s_{22}\vert$, and where $\mathbf{E}_{ij}$ is the electric field inside the structure when excited from port $j$ ($i$ being the output port). As a proof of concept, this behavior, showing a trade-off between nonreciprocal transmission ratio (NTR) and insertion loss, is discussed in Fig.~\ref{fig:1} by first assuming a nonlinear nondispersive (lossless) ENZ material with dielectric permittivity $\epsilon_{ENZ}\!=\!0.1$ which, for a given nonlinear susceptibility $\chi^{(3)}$ (we herein consider Kerr-like instantaneous third-order processes only), sees its effective nonlinear response increase with decreasing $\epsilon$ \cite{solis2020dependence}.

The origin of the above mentioned trade-off is clearly seen in panels (b1),(b2) as a function of the Si layer's thickness $D_{Si}$ ($D_{ENZ}$ is fixed to one free-space wavelength $\lambda$): in panel (b1), the solution to the linear problem under normally-incident plane-wave excitation is computed with a Transfer Matrix Method (TMM) \cite{Sipe:87} and shows how the ratio of $\mathbf{x}$-averaged electric-field magnitudes inside the ENZ layer $\frac{\vert E_{ENZ,12}\vert}{\vert E_{ENZ,21}\vert}$ (green line) finds its first maximum very close to the first minimum of transmission (black line $\vert s_{21}\vert\!=\!\vert s_{12}\vert$) at $D_{Si}\!=\!0.0695\lambda$, whereas this ratio is first equal to 1 at (besides $D_{Si}\!=\!0$) $D_{Si}\!=\!\frac{\lambda}{2\sqrt{\epsilon_{Si}}}$ ($\epsilon_{Si}\!=\!12.36$), also very close to the first maximum of transmission at $D_{Si}\!=\!0.1477\lambda$. In panel (b2), we consider the (monochromatic) nonlinear problem parameterized by $\chi^{(3)}(\omega;\omega,-\omega,\omega)\!=\!3.33\!\times\!10^{-21}\text{(V/m)}^{-2}$---i.e., only the self-phase modulation (SPM) process is taken into account \cite{Boyd:2008:NOT:1817101}---and find $s_{21}$ and $s_{12}$ with a nonlinear finite-difference frequency-domain (FDFD) method \cite{Szarvas:18} for an electric-field phasor strength of $E_0\!=10^9\text{(V/m)}$, which show an NTR of $\frac{\vert s_{12,nl}\vert}{\vert s_{21,nl}\vert}$ (green line) that is precisely maximized (minimized) around $D_{Si}\!=\!0.0695\lambda$ ($D_{Si}\!=\!0.1477\lambda$), in agreement with $\frac{\vert E_{ENZ,12}\vert}{\vert E_{ENZ,21}\vert}$ in (b1). Panels (c1),(c2) show, for these two thicknesses of interest, the electric-field profiles vs. $\mathbf{x}$ of the linear and nonlinear scenarios under forward and backward transmission. Whereas in (c1) $\vert E_{21,l}(x)\vert\!\ll\!\vert E_{12,l}(x)\vert$ and thereby the nonlinear response is perturbative in the forward direction, in (c2) $\vert E_{21,l}(x)\vert$ and $\vert E_{12,l}(x)\vert$ inside the ENZ slab are essentially the mirror image of each other and, consequently, so are $\vert E_{21,nl}(x)\vert$ and $\vert E_{12,nl}(x)\vert$, which leads to $\text{NTR}\!\approx\!1$ (the subscripts $l$ and $nl$ stand for \textit{linear} and \textit{nonlinear}, respectively). Panels (d1),(d2) further exemplify this nonreciprocal transmission when considering an instantaneous nonlinear polarization of the form $P_{nl}(z,t)\!=\!\epsilon_0\chi^{(3)}E^3(z,t)$, which also accounts for e.g. harmonic-generation processes like $(3\omega;\omega,\omega,\omega)$. We address this nonlinear polarization with a nonlinear finite-difference time-domain (FDTD) method \cite{558652} and plot the temporal profile of the resulting transmitted electric field for excitation from opposite ports for the two thicknesses of interest: whereas in (d1) $E_{12,nl}$ presents a much larger high-harmonic content than the perturbatively nonlinear output $E_{21,nl}$, in (d2) we have that $E_{12,nl}\!\approx\!E_{21,nl}$ and harmonic distortion is notable for both forward and backward incidence. (We note that we choose to neglect two-photon absorption and self-phase modulation in the silicon layer; this is justified because, as seen in the next section, the field enhancement takes place inside the ITO film. In any case, other high-index dielectrics could be used instead of silicon.)

\section{Realistic Design with ITO}
When we consider realistic materials and bring dispersion into the picture, the efficiency of the overall nonlinear response is damped by losses and, in the case of harmonic generation, by phase mismatch as well. Regardless, thin films of transparent conducting oxides (TCOs) such as Al-doped ZnO (AZO) \cite{PhysRevLett.116.233901} and indium tin oxide (ITO) \cite{Alam795} have shown, in the frequency range of vanishing real part of their dielectric function, an unprecedentedly large effective nonlinear refractive index \cite{PhysRevLett.116.233901,Alam795} and enhanced harmonic generation \cite{doi:10.1063/1.4917457,Capretti:15,doi:10.1021/acsphotonics.5b00355,Yang2019}, especially for TM-polarized light under oblique incidence maximizing the component of the electric field normal to the boundary, $E_x$ \cite{PhysRevB.87.035120}. This has been done by coupling the incident transverse wave either to the bound plasmon polariton ENZ mode---with e.g. the Kretschmann configuration as in \cite{doi:10.1063/1.4917457}---or to the leaky Ferrell–Berreman mode from the longitudinal bulk plasmon resonance \cite{Capretti:15,doi:10.1021/acsphotonics.5b00355}. Following this reasoning, we will adopt the latter approach and design a Si-ITO thin bilayer structure that simultaneously maximizes both absorption and $\vert E_x\vert$---this latter is approximately constant across the thickness of a (deeply-)subwavelength film---in the linear problem at the ENZ frequency when the ITO port is fed (ITO's permittivity is parameterized with a Drude model; see Appendix A for more details); for moderate nonlinear responses, this operational point will thus be in the vicinity of a peak of NTR. A maximum absorptance of 0.83 is found around $D_{Si}\!=\!80$ nm and $D_{ITO}\!=\!100$ nm in the usual $40^{\circ}\!-\!55^{\circ}$ range.
\begin{figure*}[t]
\includegraphics[width=7in]{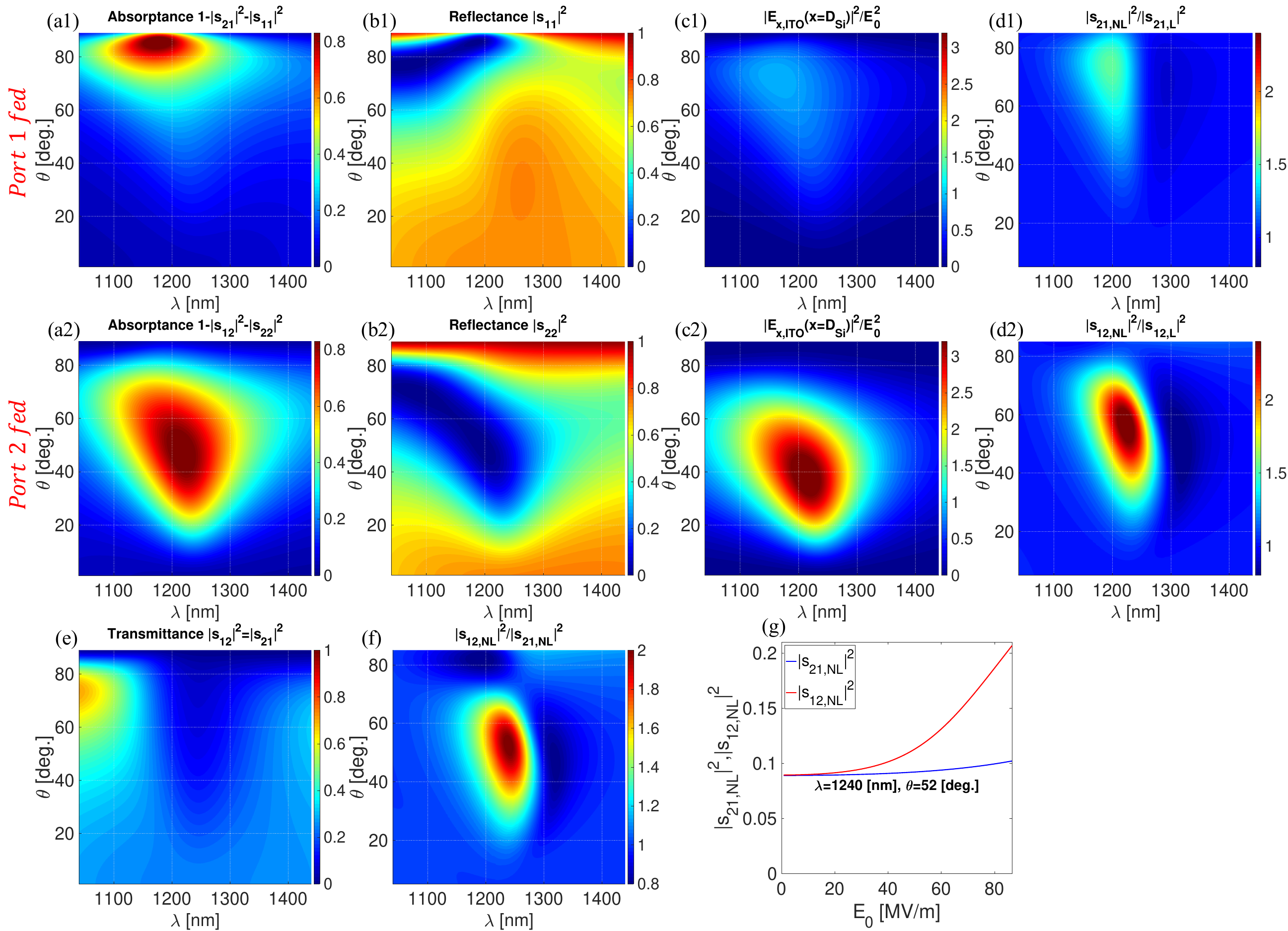}
\caption{Nonlinearity-induced nonreciprocal response vs. $(\lambda,\theta)$. (a1)-(d1) Linear absorptance, reflectance, normalized $\vert E_x(x\!=\!D_{Si})\vert$, and nonlinear-to-linear transmittance ratio, respectively, for excitation from port 1 (panels (a2)-(d2) for port 2). (e),(f) Linear (reciprocal) transmittance and squared NTR, respectively. (g) Dependence of the nonlinear nonreciprocal response vs. incident electric field for the optimal $(\lambda,\theta)$-point taken from panel (f), the optimal $\lambda$ being precisely the ENZ wavelength.}
\label{fig:3}
\end{figure*}

With these two thicknesses, we will move on to our nonlinear problem, but let us first compare, in Fig.~\ref{fig:3}, absorptance (panels (a1),(a2)) and normal component of the electric field (panels (c1),(c2)) when exciting the optimized linear structure from either side: when port 2 is excited (panels (a2),(c2)), an absorption peak of 0.86 is achieved at $(\lambda,\theta)\approx(1214\text{nm},38.5^{\circ})$, whereas the maximum enhancement of $\vert E_x\vert^2$ at the Si-ITO interface, equal to 3.39, is located at $(\lambda,\theta)\approx(1216\text{nm},47.5^{\circ})$; we will use this asymmetry in the field distribution with respect to excitation from port 1 in order to maximize the sought nonreciprocal response. Incidentally, note the slight angle shift between the peaks of absorption and $\vert E_x\vert$: these two peaks would tend to converge in a situation of perfect absorption, but such situation would in this case only be found for a virtual mode in the complex frequency (or complex transverse wavevector) plane \cite{Vassant:12}. Additionally, the reflectance maps in panels (b1),(b2) illustrate how the absorption peak more or less overlaps with a zero in the reflectivity and can thus be linked to the pseudo-Brewster angle. Restricting for now our nonlinear problem to SPM at the ENZ frequency and assuming a third-order nonlinear susceptibility of $\chi^{(3)}\!=\!6.67\!\times\!10^{-18}\text{(V/m)}^{-2}$ (Appendix C explains in detail why such value) and $E_0\!=\!8.66\!\times\!10^7\!\text{(V/m)}$, the nonlinear transmittance enhancement for both ports are mapped in panels (d1),(d2), while the squared NTR is shown in panel (f), showing a maximum $\frac{\vert s_{21,nl}\vert^2}{\vert s_{12,nl}\vert^2}$ ratio of 2.04 at the point $(\lambda,\theta)\!\approx\!(1240\text{nm},52^{\circ})$, i.e., exactly at the ENZ wavelength (note also that, for excitation from port 1, there is close to grazing incidence an absorptance peak of 0.93 at $(\lambda,\theta)\!\approx\!(1193\text{nm},86^{\circ})$ that, nonetheless, due to low $|E_x|$ enhancement, leads to a maximum $\frac{\vert s_{21,nl}\vert^2}{\vert s_{12,nl}\vert^2}$ ratio of only 1.27). Finally, we choose this point to measure the strength of the nonlinear nonreciprocal response vs. the intensity of the incident light in panel (g), where the nonlinearity in $\vert s_{12}\vert^2$ (red) is much stronger than in $\vert s_{21}\vert^2$ (blue).

These nonlinear simulations were performed with the same one-dimensional (1D) FDFD solver of Fig.~\ref{fig:1}: for plane-wave oblique incidence, we can simplify the periodic boundary conditions across $\mathbf{y}$ and shrink the nonlinear monochromatic two-dimensional (2D) problem down to a 1D grid by simply enforcing $\frac{\partial}{\partial y}\!=\!ik_{y,inc}$ ($e^{-i\omega t}$ convention). Furthermore, above the intensity level given by the chosen $E_0$, the Newton predictor-corrector iterative scheme of our nonlinear solver does not converge when $\vert E_x\vert$ is maximum at the pseudo-Brewster angle, which is perfectly consistent. In this regard, Appendix B shows how one could in principle achieve similar NTR values with dramatically less intensity by reducing the losses in ITO.

\section{Nonreciprocal Response to a Pulsed Laser Beam}
\begin{figure}[H]
\centering
\includegraphics[width=3.4in]{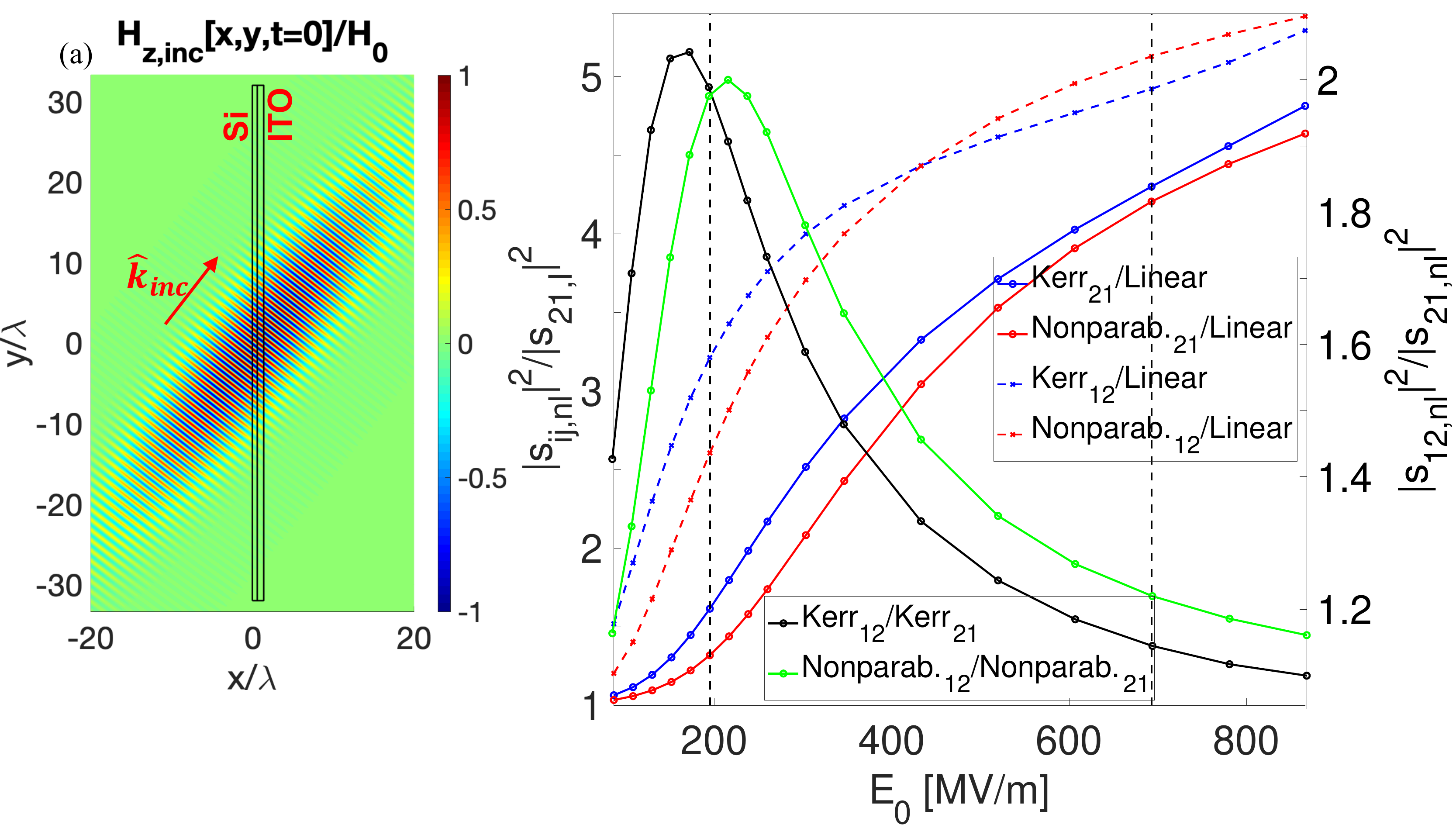}
\caption{Nonreciprocal response for excitation with ultrafast pulse. (a) Schematic drawing of the numerical experiment setup for excitation from port 1 and superimposed incident magnetic field $H_z$ when the maximum of intensity crosses the bilayer's middle point ($t\!=\!0$). (b) Nonlinearity-induced nonreciprocal response vs. incident electric field intensity (actually, its phasor halved value, for comparison with Fig.~\ref{fig:3}d), assuming both Kerr-like (fast) and thermal (slow) nonlinearities: the blue and red (black and green) curves represent nonlinear-to-linear (nonlinear backward(12)-to-forward(21)) transmittance ratios, respectively.}
\label{fig:5}
\end{figure}

In the above section we considered a realistic Drude-like dispersion for $\chi^{(1)}$ but narrowed the nonlinear problem down to SPM at a single frequency in a steady state. A more realistic approach needs to consider (i) a richer ensemble of nonlinear processes and (ii) the dispersive nature of the relevant nonlinear susceptibilities. In Appendix C we briefly go over how the root of TCOs' nonlinear processes (including ITO) lies in the nonparabolicity of their conduction band, and make the distinction between fast (anharmonic) and slow (thermal) nonlinearities \cite{Kinsey2019,https://doi.org/10.1002/lpor.202000291}: for our purposes it suffices to neglect the dispersion of the former and consider the same instantaneous Kerr-like polarization response of Sec.~II, viz. $P_{Kerr,j}(\mathbf{r},t)\!=\!\epsilon_0\chi^{(3)}E_j^3(\mathbf{r},t)$, $j\!=\!x,y$ being the cartesian component; the latter reduces the ``hot'' plasma frequency $\omega_{p,h}$ through laser-induced heat \cite{Guo2016,Alam2018}, yielding an effective nonlinear susceptibility $\chi^{(3)}_{eff}$ two orders of magnitude larger. In this respect, an adequate comparison of these two is best done in the time domain, where all frequency-mixing processes are more easily simulated in the former case, and the delay effect of the latter emerges. Moreover, high-power lasers hardly operate in a continuous-wave mode. Besides, the nonlinearity enhancement due to low group velocity at the ENZ frequency \cite{Secondo:20} is now taken into account. We will thus now perform numerical experiments by exciting our proposed nonreciprocal structure with a high-intensity ultrafast pulse (see Appendix D for details). The setup of the numerical experiment is shown in Fig.~\ref{fig:5} (panel (a)), where we compare the normalized nonlinear transmissions for both nonlinearities as a function of the laser intensity. For the sake of discussion, we are again considering, for the Kerr interactions, $\chi^{(3)}\!=\!\frac{\text{Re}[\chi^{(3)}_{eff}]}{3}$, i.e., fast and slow nonlinearities are (approximately) equally strong (see Appendix C). In panel (b), we show how the $\frac{\vert s_{21,nl}\vert^2}{\vert s_{12,nl}\vert^2}$ ratio reaches its maximum of around 2---commensurate with the results in the frequency domain of Fig.~\ref{fig:3}---for approximately $E_0\!=\!200 \text{MV/m}$ (black and green solid lines). Interestingly, although beyond this field intensity the nonlinear response keeps increasing (blue and red plots), this tends to saturate and the NTR decays. Importantly, we should mention that, as opposed to the slow nonlinearity's characterization in Appendix C, saturation effects are in no way parameterized by a single instantaneous third-order susceptibility; so one could at most state that the Kerr nonlinear transmission (solid and dashed blue)---and not the Kerr nonlinear response \emph{per se}---saturates in the range of intensities considered.

\begin{figure*}[t]
\centering
\includegraphics[width=6in]{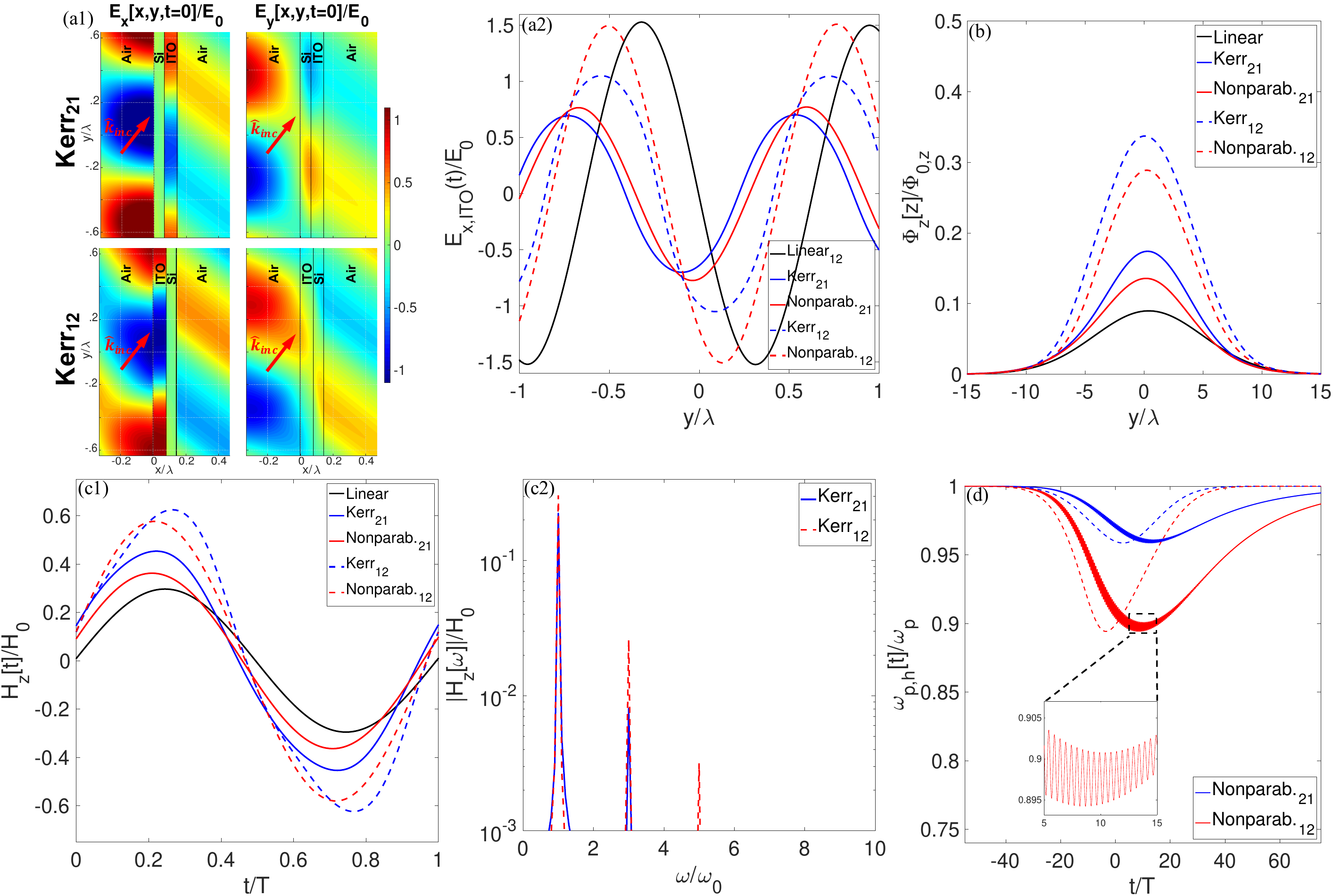}
\caption{Nonlinear response of the nonreciprocal device for maximum NTR: $E_0\!=\!195 \text{MV/m}$. (a1) $\mathbf{x}\mathbf{y}$-maps of (normalized) $E_x$ and $E_y$ at $t\!=\!0$ for excitation from both ports and Kerr nonlinearity only. (a2) 1D-cuts of $E_x$ in the middle of the ITO film (fixed $\mathbf{x}$) and $t\!=\!0$ vs. $\mathbf{y}$, but considering all linear, Kerr and thermal scenarios. (b) Normalized flux of energy density across the bilayer's output plane ($\Phi_x$ vs. $\mathbf{y}$) corresponding to the scenarios in panel (a2). (c1) Time profile of (normalized) transmitted $H_z$ at $\mathbf{y}\!=\!0$ (i.e., very close to beam axis). (c2) Spectral content of the Kerr-signals in (c1), with third and fifth harmonics showing up. (d) Time-varying nonlinear (thermal) plasma frequency at the $\mathbf{x}\mathbf{y}$-center of the ITO film (excitation from opposite inputs included); the dashed lines represent the complement of a normalized measure of the time-average intensity, defined as $1\!-\!\frac{\langle\|\mathbf{\tilde{E}}(t)\|^2\rangle_T}{max\left[\langle\|\mathbf{\tilde{E}}(t)\|^2\rangle_T\right]} \left(1\!-\!\min\left[\frac{\omega_{p,h}(t)}{\omega_p}\right]\right)$, at this same point, while the inset shows the second-harmonic ripples (see Appendix C).}
\label{fig:6}
\end{figure*}

In Fig.~\ref{fig:6} we look in detail into the case for which $E_0\!=\!195 \text{MV/m}$ (first vertical dashed line in Fig.~\ref{fig:5}). Panel (a) shows, for the fast nonlinearity, the near-field maps of $E_x$ and $E_y$ when the incident Gaussian pulse is at its maximum ($t\!=\!0$); strikingly, the enhancement of $E_x$ when excitation happens from the ITO side is somewhat quenched in the harmonic-generation process. This can be more clearly seen in panel (a2) when juxtaposing $E_x(y,t\!=\!0)$ at the $\mathbf{x}$-center of the ITO slab (dashed blue) with the linear and nonparabolic-nonlinear cases (solid black and dashed red, respectively). The results included in Appendix E for a higher light intensity (see panel (a2) in Fig.~\ref{fig:7}) replicate this behavior also for the thermal nonlinear process, confirming the intuition behind what is perhaps the key take-away message: as seen in Fig.~\ref{fig:5}b, there is a fundamental upper bound to the nonreciprocal response achievable with this structure, imposed by the simple fact that, as the nonlinear response increases/saturates, mode confinement in the direction that excites this mode---and thus the asymmetry between opposite directions---tends to vanish (the permittivity of the ITO layer is dominated by self- and cross-phase modulation at large
intensities \cite{PhysRevA.88.043812}, thus detuning the bi-layer from optimal coupling conditions); in other words, the response tends to become reciprocal for huge optical intensities. The only open route for increasing the NTR further would be by reducing loss and hence the baseline linear (reciprocal) transmission: in the limit of negligible loss, perfect absorption could in principle be achieved, which would greatly boost this NTR ratio.

Panel (b) depicts the flux of energy density across $\mathbf{x}$ at the output of the device, defined as $\Phi_x(y)\!=\!\int_{-\infty}^{\infty}\widehat{\mathbf{x}}\!\cdot\!\big(\mathbf{E}(y,t)\!\times\!\mathbf{H}(y,t)\big)dt$
and normalized by the incident flux at the axis of the Gaussian beam, $\Phi_{0,x}\!=\!\frac{\sqrt{\pi}\sigma E_0^2\cos{\theta}}{2\eta_0}$. Integrating $\Phi_x(y)$ along $\mathbf{y}$ gives us the effective $\vert s_{21}\vert^2$ and $\vert s_{12}\vert^2$ of Fig.~\ref{fig:5}b. The low-order harmonic distortion in the Kerr-like scenarios, only slightly perceived in panel (a2) (solid and dashed blue lines), is much sharper in panel (c1) representing the output $H_z$ at the $\mathbf{y}$ middle-point. A Fourier transform of the Kerr outputs in (c1), included in panel (c2), shows a third and fifth harmonic, more pronounced when the device is excited from port 2 (dashed red). Finally, panel (d) shows the time-varying $\omega_{p,h}(t)$ arising from the slow nonlinearity and its relaxation delay: the time delay between the minima of $\omega_{p,h}$ (solid lines) and the complement of the normalized electric field intensity averaged over an optical cycle $T$ (dashed lines) is around $25T$ or $\sim$100fs. Now, when $\omega_{p,h}$ is minimum, we have a maximum effective nonlinear change in the refractive index of $\Delta n\!=\!0.47\!-\!0.26i$ (in Appendix E we perform a study analogous to Fig.~\ref{fig:6}, but considering instead $E_0\!=\!692\text{MV/m}$---indicated in the second black dashed line of Fig.~\ref{fig:5}. The stronger nonlinear response is evident by inspecting the higher-harmonic content caused by the fast nonlinearity and the larger dip in $\omega_{p,h}(t)$ from the slow nonlinearity, associated with a maximum $\Delta n\!=\!0.88\!-\!0.34i$).

\section{Conclusions}
We have theoretically shown how a simple thin two-layer stack comprising highly-doped ITO and silicon provides the spatial asymmetry and, at the ENZ condition, strong nonlinear response necessary to break reciprocity. Large field confinement inside the ITO film is ultimately behind the enlarged nonreciprocal transmission and, consequently, there is an upper limit to the NTR: as the effective nonlinear response increases (be it due to intrinsically larger nonlinear susceptibility or higher incident power), this asymmetric mode confinement tends to disappear, thus killing nonreciprocity. We show how this limiting mechanism is triggered above a similar input power level, regardless of the nature of the nonlinearity (instantaneous or thermal). The only open venue left for increasing the nonreciprocal response would be to reduce loss and thus decrease the denominator in the NTR ratio, rather than increasing its numerator.

\begin{acknowledgments}
This work is supported in part by the Defense Advanced Research Projects Agency (DARPA) Defense Sciences Office (DSO) Nascent Light-Matter Interaction program under grant number W911NF-18-0369.
\end{acknowledgments}

\appendix
\section{Linear Design of the Si-ITO bi-layer: Maximization of Absorption and $\vert E_x\vert$}
\renewcommand\thefigure{\thesection.\arabic{figure}} 
\setcounter{figure}{0}
As pointed out in Sec.~III, we seek to maximize the nonlinear response, and in turn the NTR, by first achieving optimal coupling to a Berreman mode sustained by our bilayer device in the vicinity of the ENZ frequency. We do so by first performing a linear TMM-based parametric sweep of both $D_{Si}$ and $D_{ITO}$, with the pair (absorption,$\vert E_x\vert$) as objective function for maximization when the ITO side (port 2) is fed. This is shown in Fig.~\ref{fig:2}, for which a Drude–Sommerfeld model parameterizes ITO's linear dielectric function as $\epsilon(\omega)\!=\!\epsilon_{\infty}\!+\!\chi^{(1)}(\omega)$, where $\chi^{(1)}(\omega)\!=\!-\frac{\omega_p^2}{\omega^2+i\omega\Gamma}$, with free-electron linear plasma frequency $\omega_p\!=\!2.9719\!\times\!10^{15}$ [rad/s], collision rate $\Gamma\!=\!0.0468\omega_p$, and high-frequency permittivity $\epsilon_{\infty}\!=\!3.8055$, placing the ENZ condition at a wavelength of 1240 nm ($\epsilon_{ENZ}\!=\!0.35i$) \cite{Alam795}. By inspecting panels (b1),(b2) in Fig.~\ref{fig:2}, an absorptance peak of 0.83 is located at $D_{Si}\!\approx\!80$ nm and $D_{ITO}\!\approx\!100$ nm when the angle of incidence is between $40^{\circ}\!-\!55^{\circ}$.
\begin{figure*}[t!]
\centering
\includegraphics[width=6in]{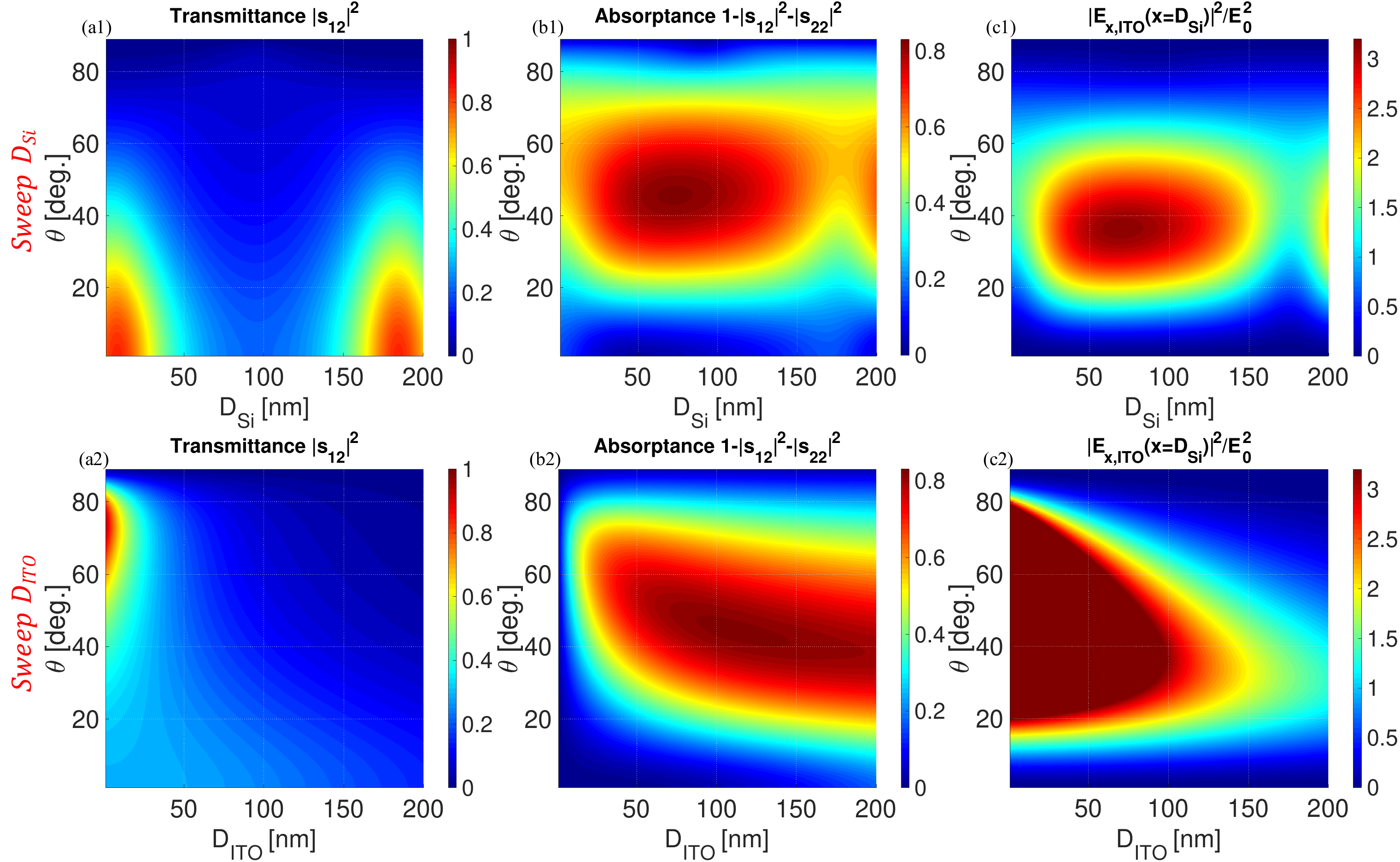}
\caption{Optimal values of Si and ITO film thickness from analysis of the linear problem for excitation from the ITO layer side (port 2). (a1)-(c1) Transmittance, absorptance and normalized (squared) longitudinal electric-field magnitude $\vert E_x(x\!=\!D_{Si})\vert$ at the ITO side of the ITO-Si interface, respectively, vs. Si film thickness $D_{Si}$ and angle of incidence $\theta$, with fixed $D_{ITO}\!=\!100$ nm. Panels (a2)-(c2) map the same quantities vs. $(D_{ITO},\theta)$, with $D_{Si}\!=\!80$ nm.}
\label{fig:2}
\end{figure*}

\section{The Effect of Loss on Efficiency}
\setcounter{figure}{0}
We repeat in Fig.~\ref{fig:4} the same calculations of Fig.~\ref{fig:3} (Sec.~III) when the collision frequency is reduced by one order of magnitude ($\Gamma\!=\!0.00468\omega_p$), which allows to achieve a similar maximum $\frac{\vert s_{21,nl}\vert^2}{\vert s_{12,nl}\vert^2}$ ratio of 2.15 at $(\lambda,\theta)\!\approx\!(1234\text{nm},22^{\circ})$, but with $E_0\!=\!6.06\!\times\!10^6\!\text{(V/m)}$, i.e., 200 times less intensity. This is consonant with the higher $Q$-factor of the new resonant device, which in the linear problem provides a backward-direction narrow absorption peak of 0.93 and a $\vert E_x\vert^2$ enhancement above 50.
\begin{figure*}[t]
\centering
\includegraphics[width=7in]{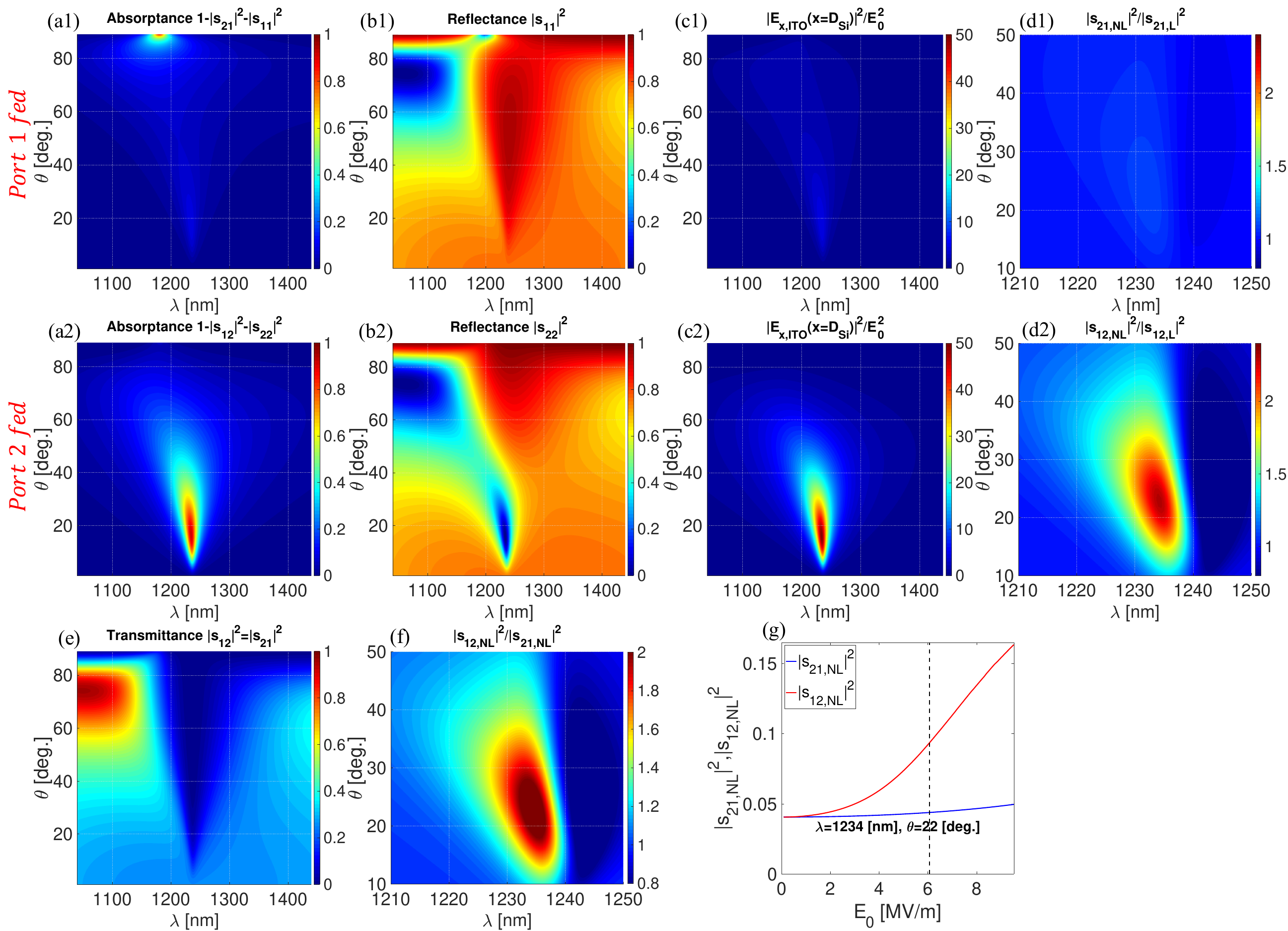}
\caption{Nonlinearity-induced nonreciprocal response vs. $(\lambda,\theta)$ if we assume ITO's collision rate $\Gamma$ is ten times smaller. The panels are the same as for Fig.~\ref{fig:3}.}
\label{fig:4}
\end{figure*}

Furthermore, Fig.~\ref{fig:9} below shows the results obtained from replicating the numerical experiment in Fig.~\ref{fig:5} (Sec.~IV) but for continuous-wave plane-wave incidence, with both the original $\Gamma$ and $\Gamma\!/\!10$. Irrespective of the nature of the nonlinearity (see Appendix C), the NTR decreases beyond a certain laser intensity, as seen in Fig.~\ref{fig:5} and, notably, not only can this NTR peak be greatly boosted by reducing $\Gamma$, but this peak is found at a lower light intensity.

\begin{figure}[h]
\centering
\includegraphics[width=3.4in]{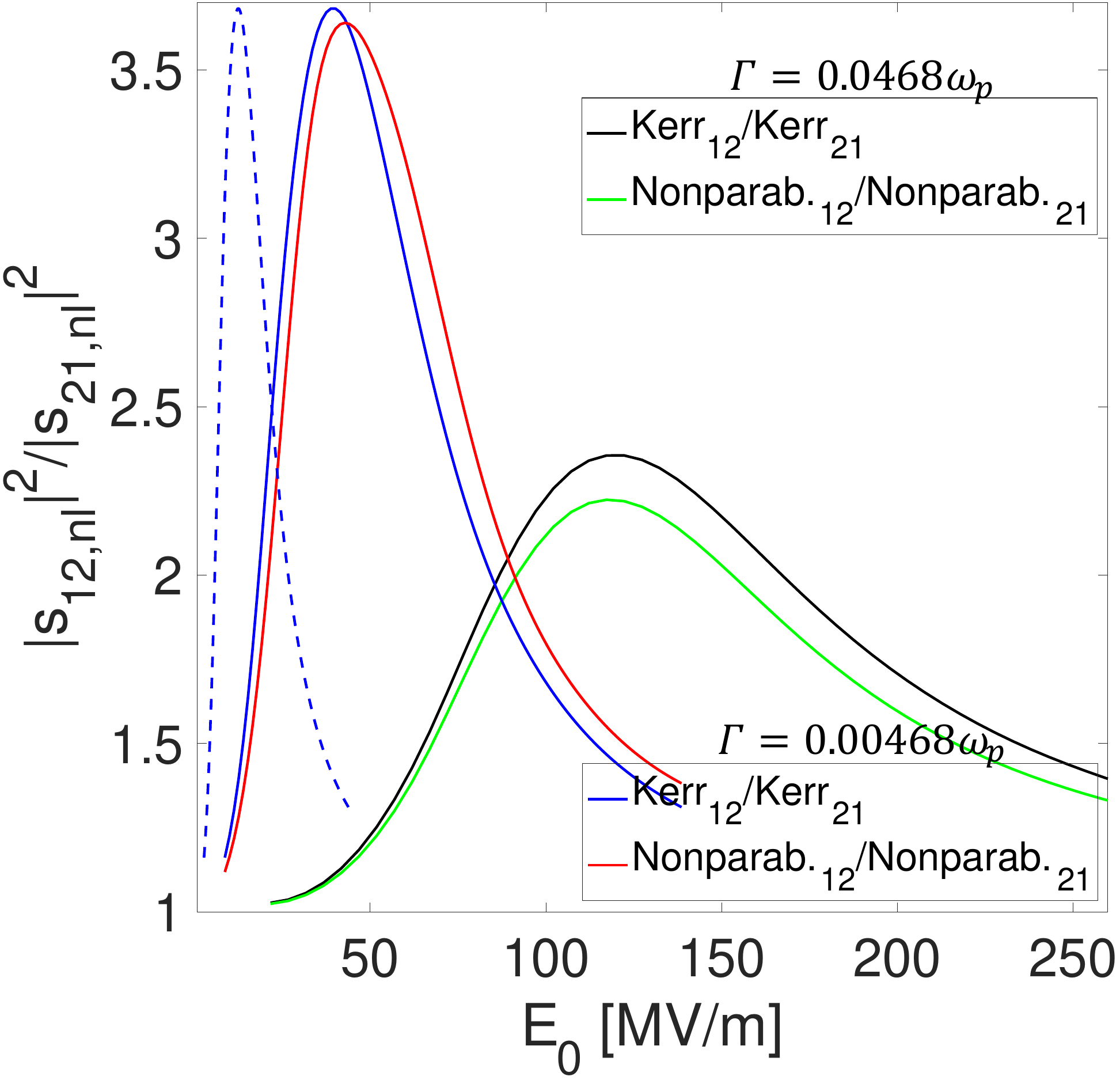}
\caption{Nonlinear transmittance ratio vs. incident electric field magnitude for $\Gamma\!=\!0.0468\omega_p$ (black and green lines) and $\Gamma\!=\!0.00468\omega_p$ (red and blue), each at its corresponding optimal point in the $(\lambda,\theta)$-plane. Both fast and slow ITO nonlinearities are considered. When $\Gamma\!=\!0.00468\omega_p$, the effective thermal susceptibility is reduced according to $\chi_{eff}^{(3)}\!\propto\!\Gamma$ (Eq.~(\ref{eq:3}), so we scale the fast susceptibility $\chi^{(3)}$ too (solid blue). If, on the contrary, we keep the initial value of $\chi^{(3)}\!=\!6.67\!\times\!10^{-18}\text{(V/m)}^{-2}$, we would have the same function compressed by a factor of $\sqrt{10}$ along the $E_0$-axis (dashed blue). These calculations are performed for steady-state plane-wave excitation in the time-domain.}
\label{fig:9}
\end{figure}

\section{Nonparabolicity of ITO's Conduction Band}
\setcounter{figure}{0}
On the one hand, there is experimental evidence of strong harmonic generation from resonant ITO thin films under oblique excitation, which can only be justified with an instant nonlinear effect, very much like anharmonic oscillations of bound electrons in dielectric materials. Certainly, ITO's free electrons are driven by the optical field within a nonparabolic energy band with energy ($\mathcal{E}$)-wavevector($k$) dispersion relation of the form $\frac{\hbar^2k^2}{2m^*}\!=\!\mathcal{E}\!+\!\mathcal{C}\mathcal{E}^2$, $\mathcal{C}$ being the nonparabolicity parameter \cite{KANE1957249} and $m^*$ the free electrons' effective mass at the bottom of the conduction band ($k\!=\!0$). The third-order susceptibility of this fast nonlinearity has been estimated to be roughly of the same order as non-resonant nonlinearities from bound electrons, around $10^{-19}\text{(V/m)}^{-2}$ \cite{https://doi.org/10.1002/lpor.202000291}.

\begin{figure}[ht!]
\centering
\includegraphics[width=3.4in]{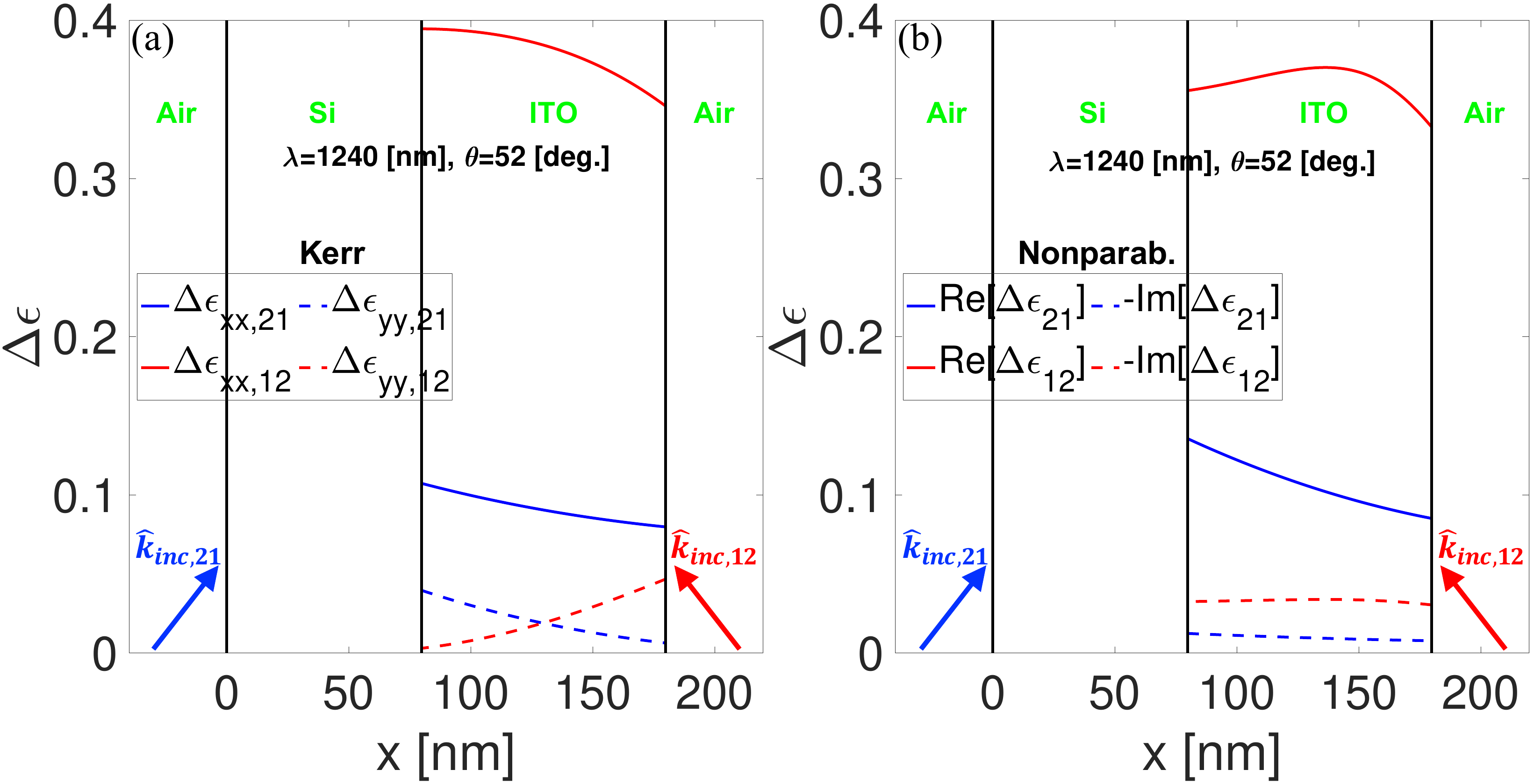}
\caption{Comparison of the effective nonlinear permittivity $\Delta\epsilon$ vs. depth into the ITO film at the optimal $(\lambda,\theta)$-point of maximum NTR in Fig.~\ref{fig:3}(f), considering both Kerr (a) and thermal (b) nonlinearities: similar values of $\Delta\epsilon$ are obtained for both types of nonlinear interactions. Incidence from port 1 (blue) and port 2 (red) are included in both panels. }
\label{fig:8}
\end{figure}

On the other hand, delayed nonlinearities involving real transitions, with much stronger response, are behind the remarkably large nonlinear refractive index of ITO observed in \cite{Alam795}. This phenomenon is also traced back to the nonparabolicity of its conduction band: the laser-induced absorption increases the temperature of free carriers, some of which are promoted above the Fermi level, thus acquiring a higher effective mass. This is explained by the fact that electron density conservation under intraband transitions lowers the electro-chemical potential which, in turn and due to the band's nonparabolicity, raises the effective mass of the sea of thermalized electrons at higher energy levels, which ultimately reduces the effective plasma frequency of our Drude model. Now, for degenerately-doped TCOs we can assume that the electrons' thermal energy is well below a Fermi level of $\sim\!1\text{eV}$ (i.e., the ratio $\kappa\!=\!\frac{U_h}{N\mathcal{E}_F}\!\ll\!1$, with $U_h$ the thermal energy density, $N$ the electron volumetric density and $\mathcal{E}_F$ the Fermi energy), which allows us to approximate the average effective mass of the electron gas as the effective mass at the Fermi level \cite{https://doi.org/10.1002/lpor.202000291} and get by without the delayed two-temperature model \cite{PhysRevB.74.024301} in describing the evolution of $U_h$ as
\begin{equation}
\frac{U_h(\mathbf{r},t)}{dt}=\mathbf{J}(\mathbf{r},t)\!\cdot\!\mathbf{E}(\mathbf{r},t)-\frac{U_h(\mathbf{r},t)}{\tau_{ep}},\label{eq:1}
\end{equation}
with the electron-phonon relaxation time $\tau_{ep}$ of only a few hundreds of femtoseconds (we will assume $\tau_{ep}\!=\!100$ fs), which explains the ultrafast response of this ``slow'' nonlinearity \cite{Kinsey:15}. The nonparabolicity-induced ``hot'' nonlinear plasma frequency can thus be substantiated as
\begin{equation}
\omega_{p,h}^2(\mathbf{r},t)\approx\frac{\omega_p^2}{1+\kappa(\mathbf{r},t)}\approx\omega_p^2\big(1-\kappa(\mathbf{r},t)\big),\label{eq:2}
\end{equation}
which intrinsically embodies the saturation effect observed in e.g. \cite{Alam795}, and where the Taylor expansion---correct to first-order---of the second equality allows for a straightforward parameterization of an effective nonlinear susceptibility in the continuous-wave scenario:
\begin{equation}
\chi^{(3)}_{eff}(\omega;\omega,-\omega,\omega)\approx-\epsilon_0\chi^{(1)}(\omega)\frac{2\Gamma\tau_{ep}\omega_p^2}{(\omega^2+\Gamma^2)N\mathcal{E}_F}.\label{eq:3}
\end{equation}
If we assume that the average effective mass $m_{eff}$ is $0.4$ times the electron mass \cite{doi:10.1063/1.4900936}, we obtain $N\!=\frac{\epsilon_0m_{eff}\omega_p^2}{q^2}\!=\!1.11\!\times\!10^{21}\text{cm}^{-3}$, which from Eq.~(\ref{eq:3}) finally yields $\chi^{(3)}_{eff}\!=\!(2.00-0.18)\!\times\!10^{-17}\text{(V/m)}^{-2}$. In Sec.~III we used $\chi^{(3)}\!=\frac{\!\text{Re}[\chi^{(3)}_{eff}]}{3}$, taking into account the degeneracy factor $3$. Rigorously speaking, however, we should note that, whereas the nonlinear term in the dielectric function is now simply a scalar $\Delta\epsilon\!=\!\chi^{(3)}_{eff}\|\mathbf{\tilde{E}}(\mathbf{r},\omega)\|^2$, in the previous section we had a second-order tensor described by $\Delta\epsilon_{xx}\!=\!3\chi^{(3)}\vert \tilde{E}_x(\mathbf{r},\omega)\vert^2$ and 
$\Delta\epsilon_{yy}\!=\!3\chi^{(3)}\vert \tilde{E}_y(\mathbf{r},\omega)\vert^2$, with $\Delta\epsilon_{yy}\!\ll\Delta\epsilon_{xx}\!\approx\!\Delta\epsilon$. This difference, illustrated in Fig.~\ref{fig:8}, barely has an effect on the nonlinear transmitted signal and the NTR maps.

Note also that only for a pulsed excitation does the delay effect of $\frac{1}{i\omega-\tau_{ep}^{-1}}$ \cite{https://doi.org/10.1002/lpor.202000291} from Eq.~(\ref{eq:1}) emerge in this thermal nonlinearity (see panel (d) in Fig.~\ref{fig:6} and Fig.~\ref{fig:7}); in a steady state, it rather manifests itself as a mere limiting factor as in Eq.~(\ref{eq:3}). Incidentally, from the dissipated power density $\mathbf{J}\!\cdot\!\mathbf{E}$ in Eq.~(\ref{eq:1}), $U_h$ not only follows the pulse intensity but also acquires a comparatively small signal oscillating at twice the carrier frequency $2\omega$ (see zoomed inset in Fig.~\ref{fig:6}d and Fig.~\ref{fig:7}d).

\section{Parameters of the Space-Time Pulsed Gaussian Beam}
For the results in Figs.~\ref{fig:5},~\ref{fig:6} (Sec.~IV), we adopt the paraxial approximation to the $\text{TEM}_{00}$ Gaussian beam mode \cite{4286012} in 2D (a beam waist radius $w_0\!=\!4\lambda$ is used; the device length in the $\mathbf{y}$-dimension is $16w_0$, enough to avoid edge effects), and modulate it with a Gaussian pulse signal $e^{-\frac{(t/\sigma)^2}{2}}$. As we choose $\sigma\!=\!75$ fs, the bandwidth of the temporal Gaussian pulse's spectrum is much smaller than the carrier frequency for near-infrared light, $\frac{1}{\sigma}\!\ll\!\omega$, and we can neglect the coupling among the beam parameters in space and time \cite{563385}.

\section{Saturation of the Nonlinear Response}
\setcounter{figure}{0}
The numerical analysis in Fig.~\ref{fig:6} (Sec.~IV) is repeated in Fig.~\ref{fig:7} by replacing the incident electric field $E_0\!=\!195 \text{MV/m}$ with $E_0\!=\!692 \text{MV/m}$. The stronger (saturated) nonlinear response comes at the expense of a large decay in the nonreciprocal response. 
\begin{figure*}[t!]
\centering
\includegraphics[width=6in]{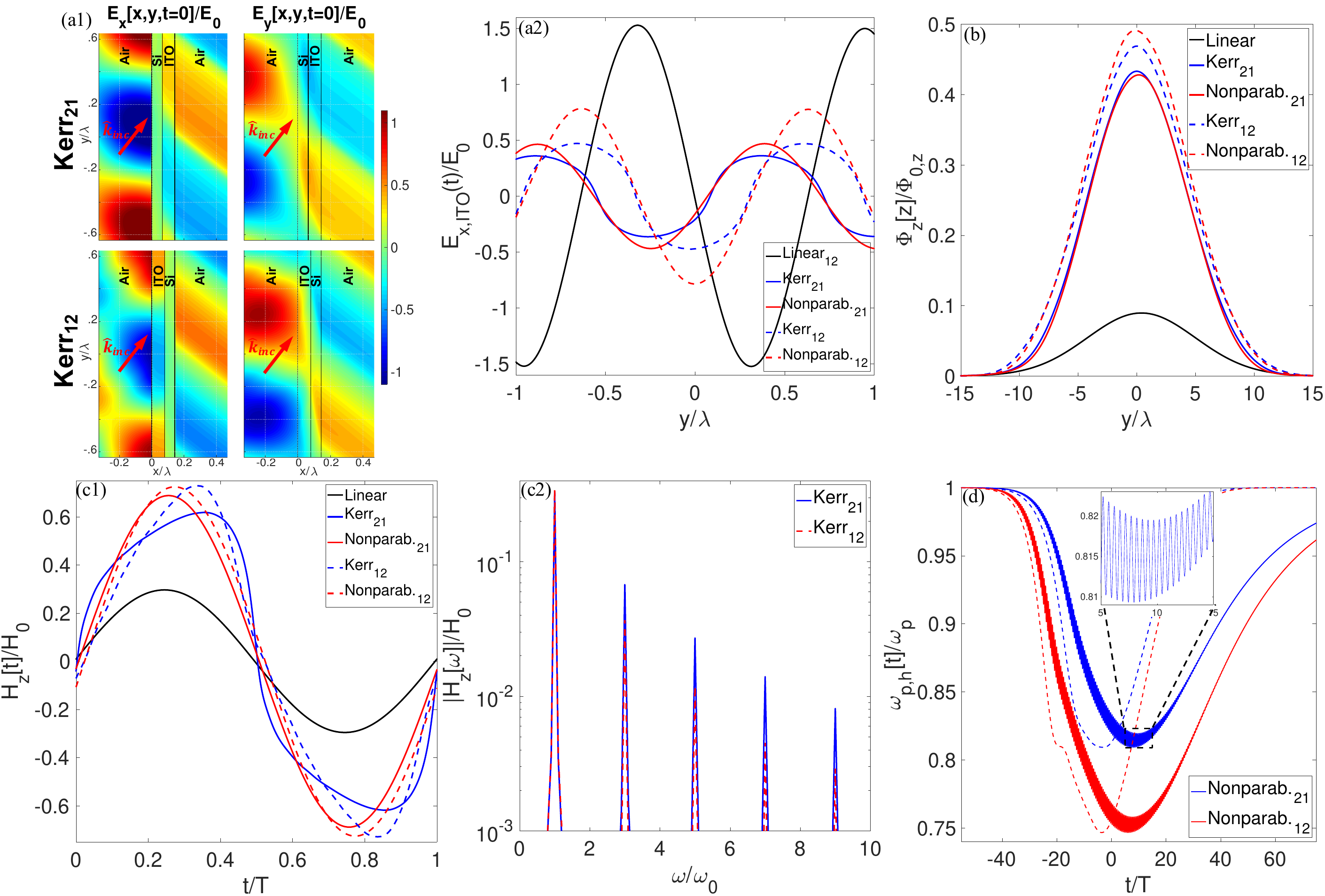}
\caption{Nonlinear response of the nonreciprocal device in the saturation regime (low NTR): $E_0\!=\!692 \text{MV/m}$. Same panels as in Fig.~\ref{fig:6} (Sec.~IV).}
\label{fig:7}
\end{figure*}

\clearpage   

\nocite{*}

\bibliography{apssamp}

\end{document}